\documentclass[a4paper,11pt,twoside]{article}

\usepackage{amsmath}
\usepackage{amssymb}
\usepackage{epsfig}
\usepackage{wrapfig}
\usepackage{array}
\usepackage{axodraw}

\makeatletter
\renewcommand{\fnum@table}{\textbf{\tablename~\thetable}}
\renewcommand{\fnum@figure}{\textbf{\figurename~\thefigure}}
\makeatother

\newlength{\myem}
\newcounter{mysubequation}[equation]

\makeatletter
\renewcommand{\section}{\@startsection{section}{1}{0em}%
        {-3.5ex \@plus -1ex \@minus -.2ex}%
        {2.3ex \@plus.2ex}%
        {\normalfont\large\bfseries}}
\renewcommand{\subsection}{\@startsection{subsection}{2}{0em}%
        {-3.25ex\@plus -1ex \@minus -.2ex}%
        {1.5ex \@plus .2ex}%
        {\normalfont\bfseries}}
\renewcommand{\subsubsection}%
        {\@startsection{subsubsection}{3}{0em}%
        {-3.25ex\@plus -1ex \@minus -.2ex}%
        {1.5ex \@plus .2ex}%
        {\normalfont\bfseries}}
\makeatother

\newcommand{\SPHT}{Service de Physique Th\'eorique, CEA-Saclay,\\ 91191 
Gif-sur-Yvette Cedex, France}
\newcommand{\UAM}{Instituto de F\'\i sica Te\'orica,
Universidad Aut\'onoma de Madrid,\\
Cantoblanco,
Spain}
\newcommand{\UCR}{Physics Department, University of California, 
Riverside,\\ California 92521, USA}

\newcommand{\preprintnumber}{}

\newcommand{\titletext}{Right-Handed Sector Leptogenesis} 
\newcommand{\authortext}{\large 
Michele 
Frigerio$^{\, a}$\footnote{E-mail address: frigerio@spht.saclay.cea.fr} , 
Thomas Hambye$^{\, b}$\footnote{E-mail address: thomas.hambye@uam.es} ~ 
and Ernest Ma$^{\, c}$\footnote{E-mail address:  ma@physics.ucr.edu} 
\medskip\\\em\normalsize 
$\mbox{}^a$ \SPHT
\\[0.1\baselineskip] 
$\mbox{}^b$ \UAM
\\[0.1\baselineskip] 
$\mbox{}^c$ \UCR}
\newcommand{\abstracttext}{Instead of creating the observed baryon asymmetry
of the universe by the decay of right-handed (RH) neutrinos 
to left-handed leptons,
we propose to generate 
it dominantly by the decay of the 
RH neutrinos to RH leptons.
This mechanism 
turns out to be successful in large regions of parameter space. 
It 
may work, in particular, at a scale 
as low as $\sim$~TeV, with no need to invoke 
quasi-degenerate RH neutrino masses 
to resonantly enhance the asymmetry.
Such a possibi\-lity can be probed experimentally by the observation at colliders 
of a 
singlet charged Higgs particle and of RH neutrinos. 
Other mechanisms which may lead to successful leptogenesis from the 
RH lepton sector interactions are also briefly presented.
The incorporation of these scenarios in 
left-right symmetric and unified models is discussed.
\flushright{SACLAY-T06/024, UCRHEP-T408}
}

\title{
\normalsize
\begin{tabular}[t]{l}\preprintnumber\end{tabular}
\vspace{1\baselineskip}\\
\Large\bfseries\titletext\bigskip}
\author{\begin{minipage}[t]{0.8\textwidth}
\normalsize\centering\authortext
\end{minipage}}
\date{}

\begin{document}
\bigskip
\maketitle
\begin{abstract}\normalsize\noindent
\abstracttext
\end{abstract}\normalsize\vspace{\baselineskip}


\section{Introduction}

In view of the recent evidence for non-vanishing neutrino masses and the belief that 
these masses are
associated with lepton number violation, the 
leptogenesis mechanism \cite{fy} has become the leading candidate to explain the 
baryon asymmetry of the universe. 
In the type I seesaw mechanism \cite{seesaw} 
the lepton asymmetry is generated
through the decay of heavy Majorana right-handed (RH) neutrinos $N_{1,2,3}$.
This scenario is naturally accommodated  in the framework of 
theories that predict the existence of RH neutrinos, 
such as $SO(10)$ \cite{georgi}, 
Pati-Salam \cite{ps} and
left-right symmetric theories \cite{ps,lr} in general. 
Another possible source of neutrino masses, which is well motivated 
in these frameworks, is the type II seesaw, which involves the interactions 
of a heavy $SU(2)_L$ triplet Higgs $\Delta_L$ \cite{mwms}.
Also this mechanism can lead to successful leptogenesis 
in agreement with the neutrino mass constraints \cite{hs,ak,sod}.
Other seesaw possibilities of successful leptogenesis arise if 
there are two or more 
heavy Higgs triplets \cite{ms,hms} or if self-conjugate  triplets of 
fermions $\Sigma$ exist \cite{hlnps}.

All the seesaw models above have the attractive feature 
that neutrino masses and baryogenesis are both generated from 
the same interactions. This nice feature has nonetheless a price to pay: due 
to the smallness of the neutrino masses, leptogenesis can be generically 
successful only 
at a very high scale \cite{bcst,th,di} (i.e. if $M_{N_1} \gtrsim 4 \cdot 10^8$
GeV \cite{di,bdp,gnrrs} or
if $M_{\Delta_L} \gtrsim 2.5 \cdot 10^{10}$ GeV \cite{hrs,hs,hms} or 
if $M_{\Sigma_1} \gtrsim 1.5 \cdot 10^{10}$ GeV \cite{hlnps}). Only 
in the special scenario where two heavy states have a quasi-degenerate mass 
spectrum, leptogenesis can be successful at much lower scales, thanks to the 
resonant 
enhancement of the lepton asymmetry occurring in this case.
Not considering this last possibility, beside the fact that these bounds are 
in tension with the gravitino constraint in 
supergravity theories, basically they imply 
that leptogenesis could {\it never} be tested directly.\footnote{
Moreover, the lower bounds on seesaw particle masses are saturated for specific structures of the neutrino Yukawa 
couplings, while generic structures require higher values of the masses. 
In particular, if the neutrino Yukawa couplings are 
analog to the charged fermion ones (as in most unified models), 
the produced lepton asymmetry in type I models is in general too 
small \cite{YH}.
In this case, even leptogenesis at high scale is successful 
only if a resonant enhancement of the asymmetry occurs \cite{AFS}.
}

In this connection, one may ask if the minimal theories incorporating 
the seesaw mechanism contain other sources of lepton asymmetry, which 
are not suppressed by the smallness of neutrino masses and, in this case, if
leptogenesis could be successful  
at a lower scale. More generally it is phenomenologically interesting
to determine what are the basic mechanisms which can induce successful 
leptogenesis at the low scale (independently of specific grand-unified 
realizations). Note that this 
does not necessarily require two sources of lepton number violation, one 
for neutrino masses 
and a different one for leptogenesis. In fact, in this paper 
we consider the case where the source of lepton number 
violation remains the same  
(i.e. the  Majorana masses of RH neutrinos $N_i$), but where the 
interactions at the origin of the decays,
instead of involving the left-handed Standard Model (SM) leptons, involve 
the right-handed SM leptons.

We consider a basic mechanism where the $N_i$'s decay to a 
charged RH lepton and a scalar charged $SU(2)_L$ 
singlet $\delta^+$ (section 2). The case where the $\delta^+$ is accompanied by a $\delta^0$ and a $\delta^{++}$ to form an $SU(2)_R$ triplet
is similar and discussed in section 3.
The issue of the incorporation of these two basic 
leptogenesis mechanisms in left-right symmetric and/or unified models
is discussed in section 4. In section 5 we identify other possible 
sources of lepton asymmetry involving RH leptons. In section 6 the 
perspectives to observe a $\delta^+$ (and a RH neutrino) with mass 
of the order of TeV are briefly outlined.

Let us notice that there is a very small number of possibilities to generate
leptogenesis at the low scale from two-body decays involving the SM 
fermions in the final state. 
In particular, the decaying particle has to be a SM gauge singlet in order to avoid very large washout from gauge scattering at low scale.
With seesaw interactions to generate the neutrino masses,
the mechanism we consider in sections \ref{singlet} and \ref{triplet}
appears to be the most 
economical non-resonant one (in terms of particle content 
and assumptions). 
Other non-resonant possibilities of inducing low scale 
thermal leptogenesis arise if the neutrino masses are not induced by seesaw 
interactions but radiatively (from 3-body decays \cite{th} or 
from $L$ violating 
soft terms in the seesaw extended MSSM \cite{bhs})
or by considering more than 3 generations of right- and left-handed 
neutrinos \cite{aal}. Resonant possibilities have been 
considered e.g.~in \cite{fps}.


\section{The simplest model: a charged $SU(2)_L$ singlet scalar  \label{singlet}}

We first consider the minimal case 
where in addition to the SM particles there exist two or more RH neutrinos $N_i$ 
and a charged scalar $SU(2)_L$ singlet $\delta^+$. In full generality 
the relevant interactions are:
\begin{eqnarray}
{\cal L} &\owns& -M^2_\delta \delta^{+\dagger} \delta^+
+\left[-\frac{1}{2} M_{N_{i}} N^{T}_{iR} C N_{iR} 
- H^\dagger \bar{N}_{iR}  (Y_N)_{ij} \psi_{jL} \nonumber \right.\\
& &  - (Y_R)_{ij} N_{iR}^T C 
\delta^+ l_{jR} -(Y_L)_{ij} \psi_{iL}^T C i \tau_2 \delta^+ \psi_{jL}
+ {\rm h.c.}\biggr]\,,
\label{Lminimal}
\end{eqnarray} 
with $\psi_{iL}= (\nu_{iL}~l_{iL})^T$ and  
$H=(H^0~H^-)^T$.

\begin{figure}[t]
\begin{center}
\begin{picture}(200,80)(0,0)
\Line(0,50)(30,50)
\DashArrowLine(60,80)(90,80){5}
\ArrowLine(60,20)(90,20)
\Line(60,80)(60,20)
\ArrowLine(60,80)(30,50)
\DashArrowLine(60,20)(30,50){5}
\Text(5,42)[]{$N_1$}
\Text(35,32)[]{$\delta^+$}
\Text(37,71)[]{$ l_{Rk}$}
\Text(69,50)[]{$N_j$}
\Text(85,72)[]{$\delta^+$}
\Text(85,28)[]{$ l_{Ri}$}
\Text(45,3)[]{(a)}
\Line(120,50)(140,50)
\DashArrowArcn(155,50)(15,0,-180){5}
\ArrowArc(155,50)(15,0,180)
\Line(170,50)(190,50)
\DashArrowLine(190,50)(220,80){5}
\ArrowLine(190,50)(220,20)
\Text(125,42)[]{$N_1$}
\Text(155,26)[]{$\delta^+$}
\Text(156,75)[]{$ l_{Rk}$}
\Text(180,42)[]{$N_j$}
\Text(207,75)[]{$\delta^+$}
\Text(207,24)[]{$ l_{Ri}$}
\Text(170,3)[]{(b)}
\end{picture}
\end{center}
\caption{One-loop diagrams contributing to the lepton asymmetry
in the $N_1$ decay.}
\label{fig}
\end{figure}
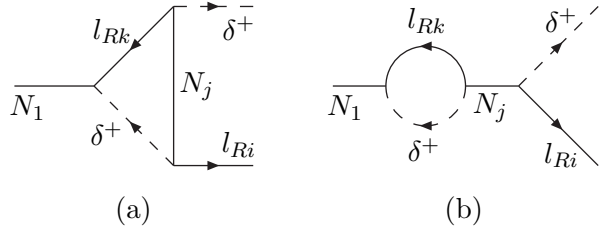

We consider the possibility that the scalar singlet is lighter than 
the RH neutrinos and we neglect, at this stage, the 
effects of the $Y_N$ couplings, which are not relevant to 
achieve our main results. 
Their effect will be quantified later.
In this case leptogenesis can 
be induced by replacing in the standard diagrams, both in the 
loop and in the 
final state, the left-handed lepton doublet with the RH charged 
lepton and the Higgs doublet with the scalar singlet, as shown in Fig.~1.
For the lightest RH neutrino $N_1$, the relevant CP asymmetry is:
\begin{equation}
\varepsilon_{N_1}=\sum_i {\frac{\Gamma (N_1 \rightarrow l_{iR} + \delta^+) - 
\Gamma (N_1 \rightarrow 
\bar{l}_{iR} + \delta^-)}{ \Gamma_{N_1}}} \cdot C_L\,,
\label{epsNdef}
\end{equation}     
with
\begin{equation}
\Gamma_{N_1}= \frac{1}{16 \pi} M_{N_1} \sum_i |(Y_R)_{1i}|^2 \,.
\label{width1}\end{equation}
In Eq.~(\ref{epsNdef}), $C_L$ is the lepton number produced in the 
decay $N_1\rightarrow l_{iR}+\delta^+$. 
Unlike the Higgs doublet in the standard 
leptogenesis case, $\delta^+$ does not have a vanishing lepton number. Once 
produced from the decay of the RH 
neutrinos,  it decays
to 2 left-handed antileptons, via the $Y_L$ couplings, so that it has $L=-2$ 
which gives $C_L=-1$. One finds
\begin{equation}
\varepsilon_{N_1}=\frac{1}{8 \pi} C_L \sum_j
\frac{{\cal I}m[(Y_R 
Y_R^\dagger)_{1j}^2]}{\sum_i |(Y_R)_{1i}|^2
}\sqrt{x_j}
\left[1-(1+x_j) \log\left(1+\frac{1}{x_j}\right)+\frac{1}{2}\frac{1}{1-x_j}\right] \,,
\label{epsN}
\end{equation}
where $x_j=M^2_{N_j}/M^2_{N_1}$. For this calculation we 
neglected $(M_\delta/ M_{N_1})^2$ corrections which are small as soon as 
the $\delta^+$ is a few times lighter than $N_1$ as we assume here.
In the limit where we also neglect the $M^2_{N_1}/M^2_{N_{2,3}}$ 
corrections, we get
\begin{equation}
\varepsilon_{N_1}=- C_L \frac{1}{8 \pi} \sum_j
\frac{{\cal I}m[(Y_R 
Y_R^\dagger)_{1j}^2]}{\sum_i |(Y_R)_{1i}|^2
}\frac{M_{N_1}}{M_{N_j}}
 \,.
\label{epsNhierarc}
\end{equation}
Apart for the $C_L$ factor and 
for a combinatoric factor of two in the self-energy contribution, 
this asymmetry is the same as in 
the standard case, replacing the ordinary Yukawa couplings $Y_N$ by the
$Y_R$ scalar singlet ones. 
Contrary to the standard case, however, the RH Yukawa 
couplings $Y_R$ do not induce 
any neutrino masses and so are not constrained by them. As a result
this mechanism may easily lead to successful leptogenesis and may also 
work at a much lower scale,
as explained below, which is phenomenologically interesting.

Considering for simplicity only 2 RH neutrinos $N_{1,2}$ (the 
effect of $N_3$ can be straightforwardly incorporated), numerically the 
constraints for successful leptogenesis are the following:
\begin{itemize}

\item The total baryon asymmetry produced is given by:
\begin{equation}
\frac{n_B}{s}=-\frac{28}{79} \frac{n_L}{s}=
-\frac{135 \zeta(3)}{4 \pi^4 g_\star}\frac{28}{79}\varepsilon_{N_1} \eta =
-1.36 \cdot 10^{-3}\varepsilon_{N_1} \eta \,,
\label{asymconstr}
\end{equation}
where $\eta$ is the efficiency factor and $g_\star=108.75$. For a maximal 
efficiency, $\eta=1$, the requirement
to reproduce the data (i.e.~$\frac{n_B}{s}=9 \cdot 10^{-11}$) implies
that 
\begin{equation}
Y_R^{(2)}\equiv \sqrt{\left| \frac{{\cal I}m \left[\sum_i(Y_R)_{1i}(Y_R^*)_{2i}\right]^2}
{\sum_i (Y_R)_{1i}(Y_R^*)_{1i}}\right| } \ge 1.3 \cdot 10^{-3} 
\sqrt{\frac{M_{N_2}}{M_{N_1}}}~,
\label{up2}\end{equation}
which means that
at least one of the $(Y_{R})_{2i}$ coupling is 
of order $10^{-3}\cdot\sqrt{ M_{N_{2}}/M_{N_1}}$ or larger.

\item To avoid washout from the $N_1$ inverse decays the constraint on the 
decay width  reads:
\begin{equation}
\Gamma_{N_1}
< H(T)|_{T=M_{N_1}}= \sqrt{\frac{4 \pi^3 g_\star}{45}} 
\frac{T^2}{M_{Planck}}\Big|_{T=M_{N_1}}\,.
\label{ooecN}
\end{equation}
Using Eq.~(\ref{width1}), the corresponding upper 
bound on the $(Y_R)_{1i}$ couplings reads
\begin{equation}
{Y}_R^{(1)} \equiv \sqrt{\sum_i |(Y_R)_{1i}|^2} < 3 \cdot 10^{-4}
\sqrt{\frac{M_{N_1}}{10^9{\rm GeV}}} ~.
\label{up1}\end{equation}
Larger values of ${Y}_R^{(1)}$
lead to suppression of the efficiency which, for successful leptogenesis, 
has to be compensated by 
larger values of the $(Y_{R})_{2i}$ couplings in the 
numerator of the asymmetry $\epsilon_{N_1}$.

\item If Eq.~(\ref{ooecN}) is satisfied, the washout from $\Delta L=2$ 
scattering mediated by $N_1$ is negligible, see e.g.~\cite{bdp}. 
Taking 
values for $(Y_{R})_{2i}$ consistent with Eq.~(\ref{up2}), the 
washout from $\Delta L = 2$ 
scatterings mediated 
by $N_{2}$ is generically negligible, except  
possibly for $M_{N_1}$ as low as a few TeV. In fact, this depends 
on the interplay 
of $Y_R^{(1)}$, $M_{N_1}$, $M_{N_2}/M_{N_1}$ as well 
as of the $(Y_{R})_{2i}$ couplings.
Large $(Y_{R})_{2i}$ couplings lead to a large CP asymmetry but also to
large $\Delta L = 2$ washout. Large $M_{N_2}/M_{N_1}$ leads to 
suppressed washout but also to a small CP asymmetry. Small
$Y_R^{(1)}$ leads to late $N_1$ decay, and therefore to 
suppressed $N_2$ washout at the 
temperature of the decay, i.e.~to a large Boltzmann suppression
of the on-shell $N_2$ contribution and to a large suppression 
of the off-shell $N_2$ scatterings (through $T/M_{N_2}$ powers and also 
a Boltzmann factor below the
$M_\delta + m_{l}$ threshold). 
The interplay of all these effects can be determined from
the Boltzmann equations. Considering them explicitly, we have checked that
even at scales as low as a few TeV, an efficiency of 
order one can be obtained easily (see also \cite{th,bhs,pil2}).

\end{itemize}

Combining the 3 constraints above, successful leptogenesis can be 
achieved in a large region of parameter space.
The scale at which the lepton asymmetry may be produced depends 
on the hierarchy between the $Y_R$ couplings of $N_2$ and $N_1$. 
This can be quantified by combining Eqs.~(\ref{up2}) and (\ref{up1}):
\begin{equation}
\frac{Y_R^{(1)}}{Y_R^{(2)}} < 0.2 \cdot  \sqrt{\frac{M_{N_1}}{M_{N_2}}\frac{M_{N_1}}
{10^9{\rm GeV}}} ~.
\end{equation}
This condition is easily satisfied for $M_{N_1} \simeq 10^{9-15}$ GeV.
When, for example, $M_{N_1}/M_{N_2}\sim 0.1$ 
and $M_{N_1}=10^7$ GeV, 
at least one of the $(Y_R)_{2i}$ couplings needs to 
be about two orders of magnitude larger than the $(Y_R)_{1i}$.
At scale as low as 1-10 TeV the hierarchy needed is more 
substantial, of about 4 
orders of magnitude, but this is not  
unrealistic for Yukawa couplings (the hierarchy 
needed is of the order of the one 
in the SM Yukawa couplings). An example of a set of parameters leading to 
an efficiency of order one
and to a baryon asymmetry in agreement with the observed one is: 
$M_{N_1}=2$ TeV, $M_{N_2}=6$ TeV, 
$(Y_{R})^{max}_{2i}\simeq 4 \cdot 10^{-3}$, $Y_R^{(1)}\simeq 10^{-7}$ 
and $M_{\delta}\simeq 750$~GeV. We find that successful leptogenesis can 
be generated with $M_{N_1}$ as low as $\simeq 1$~TeV and 
with $M_{N_2}$ as low as $\simeq 4$~TeV.\footnote{If there is an 
additional resonance 
effect, $M_{N_2}$ ($\simeq M_{N_1}$) can be lowered 
down to $\sim 1$~TeV as well.}

So far we have neglected the effects of the ordinary $Y_N$ Yukawa couplings.
Switching them on leads to more tree-level and one-loop diagrams.
In addition to the usual pure $Y_N$ diagrams there are self-energy 
diagrams involving both $Y_R$ and $Y_N$ couplings.
This leads to the full asymmetry $\varepsilon_{N_1}=\varepsilon_{N_1}^V
+\varepsilon_{N_1}^S$ where the vertex and self-energy contributions are:
\begin{equation}
\varepsilon_{N_1}^V=\frac{1}{8 \pi} \sum_j
\frac{{\cal I}m \left[C_L(Y_R 
Y_R^\dagger)_{1j}^2+2(Y_N 
Y_N^\dagger)_{1j}^2\right]}{\sum_i |(Y_R)_{1i}|^2
+ 2\sum_i |(Y_N)_{1i}|^2 
}\sqrt{x_j}
\left[1-(1+x_j) \log\left(1+\frac{1}{x_j}\right)\right] \,, 
\label{epsNV}
\end{equation}
\begin{equation}
\varepsilon_{N_1}^S=\frac{1}{16 \pi} \sum_j
\frac
{{\cal I}m \left[C_L(Y_R Y_R^\dagger)_{1j}^2
+2(C_L+1)(Y_R Y_R^\dagger)_{1j}(Y_N Y_N^\dagger)_{1j}
+4(Y_N Y_N^\dagger)_{1j}^2 \right]}
{\sum_i |(Y_R)_{1i}|^2
+ 2\sum_i |(Y_N)_{1i}|^2 
}\frac{\sqrt{x_j}}{1-x_j} \,.
\label{epsNS}
\end{equation}
As it is well-known, at scales above $\simeq 4 \cdot 10^8$~GeV, the $Y_N$ 
couplings can lead to successful leptogenesis and may dominate 
the asymmetry of Eqs.~(\ref{epsNV}) and (\ref{epsNS}).
At a lower scale, the light neutrino mass constraints 
generically require that the $Y_N$ couplings are smaller than 
$10^{-3}$, barring cancellations between 
different $Y_N$ entries.\footnote{This follows from the seesaw formula \cite{seesaw},
$m_\nu=-v^2 Y_N^T M_N^{-1} Y_N$, where $m_\nu$ is the mass matrix of light neutrinos and $M_N=diag(M_{N_1},M_{N_2},M_{N_3})$.}
Therefore the asymmetry of Eqs.~(\ref{epsNV}) and (\ref{epsNS})
can lead to successful leptogenesis only from large enough $Y_R$ 
couplings of $N_2$ (and/or $N_3$) as explained above. In this case the 
$Y_{N}$ couplings to $N_2$ and $N_3$ have a negligible 
effect in the numerator of Eqs.~(\ref{epsNV}) and (\ref{epsNS}), but still 
the $Y_N$ couplings of 
$N_1$ may have a significant effect, in particular 
from their contribution to the tree level decay width of $N_1$:
\begin{equation}
\Gamma_{N_1}= \frac{1}{16 \pi} M_{N_1} \sum_i |(Y_R)_{1i}|^2 
+ \frac{1}{8 \pi} M_{N_1} \sum_i |(Y_N)_{1i}|^2
\label{GammaN1tot}\,.
\end{equation}
Just as in the standard leptogenesis mechanism, there will be no inverse 
decay washout effect if $N_1$ contributes to light neutrino masses by less 
than $10^{-3}$ eV, that is to say if the solar and atmospheric mass splittings are
dominated by the contributions of $N_2$ and $N_3$. In fact,  
Eq.~(\ref{ooecN}) now implies the constraint (\ref{up1}) as well as
\begin{equation}
\frac{v^2 \sum_i |(Y_N)_{1i}|^2}{M_{N_1}} < 10^{-3} {\rm eV}~.
\label{upN1}\end{equation}
In the opposite case, larger $Y_R$ couplings to $N_2$ and/or $N_3$ are 
required for
successful leptogenesis, in order to increase $\epsilon_{N_1}$ thus 
compensating for the washout factor $\eta < 1$.

Note finally that, as we have assumed only 
one Higgs doublet $H$, the scalar singlet $\delta^+$ has no coupling 
bilinear in $H$.
For the case where there would be more than one Higgs doublet, $\delta^+$ 
can couple antisymmetrically to two different $H_i$ (just as 
in the Zee model \cite{zee}). 
Such a coupling would be dangerous because, combined with the
$Y_L$ coupling, it could induce a fast $\Delta L = 2$ scattering.
As a consequence, the basic mechanism above 
works safely if there is only one 
Higgs doublet lighter than $M_{N_1}$.
In the opposite case, leptogenesis may still work but only if these $\delta^+$ 
couplings to two different  Higgs doublets are forbidden or suppressed, or if 
instead the $Y_L$ couplings 
are forbidden or suppressed. Note that in the later situation 
the $\delta^+$ has to be considered as having $L=0$ so 
that in all equations above $C_L=1$.
Models with or without extra Higgs doublets and mechanisms to suppress 
dangerous $\delta^+$ couplings are discussed in section \ref{MB1}.


\section{The right-handed scalar triplet case \label{triplet}}

If the theory of particle interactions beyond the Standard Model contains left-right symmetry, the gauge symmetry has to be extended to include the group $SU(2)_L\times
SU(2)_R\times U(1)_{B-L}$.
In order to realize leptogenesis in this framework, the role of the 
charged singlet $\delta^+$ may be played by the charge-one component 
of an $SU(2)_R$ triplet $\Delta_R$. In this case the leptogenesis mechanism 
discussed in section \ref{singlet} is slightly modified. 
The relevant interactions are:
\begin{eqnarray}
{\cal L} &\owns& 
-M^2_\Delta Tr\Delta_R^\dagger \Delta_R 
+ \left[ -\frac{1}{2} M_{N_{i}} N^{T}_{Ri} C N_{Ri} 
- H^\dagger \bar{N}_{Ri}  (Y_N)_{ij} \psi_{jL} \right. \nonumber \\
& & - (Y_\Delta)_{ij} \psi_{iR}^T C i \tau_2 
\Delta_R \psi_{jR} + {\rm h.c.} \biggr]\,,
\label{lagr}
\end{eqnarray} 
with $\psi_{iL}= (\nu_{iL}$~$l_{iL})^T$, 
$\psi_{iR}= (N_{i}$~$l_{iR})^T$, $H=(H^0~H^-)^T$ and 
\begin{equation}
\Delta_R=
\begin{pmatrix}
\frac{1}{\sqrt{2}}\delta^+ & \delta^{++}  \\
\delta^0 & - \frac{1}{\sqrt{2}} \delta^+ 
\end{pmatrix} \,.
\end{equation}
Here we assume that $\delta^0$ has zero vacuum expectation value, that 
is, its contribution to RH neutrino masses is already 
reabsorbed in $M_{N_i}$ (see also section \ref{MB2}).

The diagrams in Fig.~1, in this case, lead to the same asymmetry as in Eq.~(\ref{epsN}), 
and successful leptogenesis leads to the same constraints,
replacing everywhere the $(Y_R)_{ij}$ couplings by $\sqrt{2}(Y_\Delta)_{ij}$.
In addition, as there is no coupling of the $\Delta_R$ to two 
left-handed leptons,
the $\delta^+$ does not have $L=-2$ as above and $C_L$ is modified.
Since we assume that the $\delta^+$ is lighter than the RH 
neutrinos, the $\delta^+$ cannot decay to two particles but instead to three,
either to a Higgs doublet and 
a lepton-antilepton pair which have $L= 0$, or into a 
Higgs doublet and a pair of antileptons which have $L= -2$. Summing 
over flavors of final-state leptons as well as of the virtual RH 
neutrinos $N_i$, we get
\begin{equation}
\Gamma(\delta^+\rightarrow l_L^+ l_L^- H^+)
=\frac{1}{192 \pi^3} \frac{M_\delta^3}{4}
\sum_{ij} \frac{(Y_{\Delta}Y_\Delta^\dag)_{ij}(Y_N Y_N^\dag)_{ji}}{M_{N_i}M_{N_j}} 
\label{deltawidth1}
\end{equation}
and
\begin{equation}
\Gamma(\delta^+\rightarrow l_L^+ l_R^+ H^-)
=\frac{1}{192 \pi^3} \frac{M_\delta^5}{16}
\sum_{ij} \frac{(Y_{\Delta}Y_\Delta^\dag)_{ij}(Y_N Y_N^\dag)_{ij}}{M^2_{N_i}M^2_{N_j}}.
\label{deltawidth2} 
\end{equation}
Since the second decay mode is suppressed by two 
extra powers of $M_\delta/M_{N_i}$, the first one is dominant,
so that $\delta^+$ has $L \simeq 0$ and $C_L \simeq +1$.
Note that 
it may be
unnecessary to know 
what is the lepton number of $\delta^+$ to determine that the 
value $C_L=+1$ must be taken 
in the CP asymmetry.
The reason is that Eq.~(\ref{deltawidth1}) (and {\it a fortiori} Eq.~(\ref{deltawidth2}))
may lead generically to a $\delta^+$ decay lifetime larger than the age of the 
Universe $t$ at the electroweak scale
(i.e.~at $T \simeq 150$ GeV \cite{bls} with $1/t \approx 2H$ for a 
radiation dominated universe):
\begin{equation}
\Gamma_{\delta^+} 
< \sqrt{\frac{16 \pi^3 g_\star}{45}} 
\frac{T^2}{M_{Planck}}\Bigg|_{T\simeq 150{\rm GeV}}\,.
\label{LD}\end{equation}
Assuming realistic values of the $Y_N$ couplings 
from solar and atmospheric data, taking for the $Y_\Delta$ 
couplings the values 
necessary for successful leptogenesis estimated in section \ref{singlet} and 
taking $M_\delta$ at least a few times smaller than $M_{N_1}$, one finds that 
Eq.~(\ref{LD}) is satisfied  for $M_{N_1}$ below $\sim 10^7$ GeV. 

The presence of the two extra states 
$\delta^{++}$ and $\delta^0$ does not play any 
significant role for leptogenesis. 
The $\delta^{++}$ couples only to 
two RH charged leptons and does not bring any source 
of L-violation, it has $L=-2$ lepton number.\footnote{At  
most, if the $\delta^{++}$ is still active at $T \sim M_{N_{1}}$, 
it will change the produced baryon asymmetry by about one percent
changing the number of active degrees of freedom by about one percent.}
The $\delta^0$ component couples either to $N_1 N_i$
with suppressed couplings (see Eq.~(\ref{up1}) with $Y_R$ 
replaced by $Y_\Delta$) 
or to two $N_{2,3}$, which have masses
above the temperature of the production of the asymmetry.

In summary, leptogenesis works in the same 
way in the case of Eq.~(\ref{Lminimal}), by means of a 
charged scalar singlet, and in the case of Eq.~(\ref{lagr}), by means of a 
RH triplet scalar, provided we replace everywhere $Y_R$ 
by $\sqrt{2}Y_\Delta$ and $C_L=-1$ by $C_L=+1$.


\section{Incorporating right-handed leptogenesis in unified gauge theories}

\subsection{The case of a charged singlet  $\delta^+$ \label{MB1}}

Let us consider the embedding of an $SU(2)_L$ singlet Higgs boson with charge one
in the simplest gauge extensions of the Standard Model.

In the presence of left-right symmetry, the RH leptons 
$\psi_{R}=(N~l_R)^T$
have quantum numbers $(1,2,-1)$ under the gauge 
group $SU(2)_L\times SU(2)_R \times
U(1)_{B-L}$. Since $(1,2,-1)\times(1,2,-1)=(1,1,-2)_a+(1,3,-2)_s$, a scalar
$\delta^+ \sim (1,1,2)$ has (antisymmetric) Yukawa couplings to 
RH leptons. In the
same way, it also couples to the left-handed leptons $\psi_L\sim (2,1,-1)$.
In these models, other Higgs bosons in addition to $\delta^+$ are needed in order to break spontaneously
$SU(2)_R\times U(1)_{B-L} \rightarrow U(1)_Y$ as well as to give Majorana masses to
RH neutrinos.

If the minimal left-right group is further extended to a Pati-Salam model,
$\delta^+$ is  accommodated into a $(1,1,10)$-multiplet under $SU(2)_L \times
SU(2)_R \times SU(4)_c$, which couples bilinearly to RH 
fermions $\sim
(1,2,\overline{4})$. The Pati-Salam group may be naturally embedded in unified
models based on $SO(10)$, with all fermions in a $16$-dimensional spinor
representation. In this case $\delta^+$ is part of a $120$ Higgs multiplet, which
has renormalizable Yukawa couplings to fermions.

Alternatively, one can consider the $SU(5)$ option for gauge coupling unification.
In this case, leptons are assigned as follows to $SU(5)$ representations: $\psi_L
\in \overline{5}_f$, $l_R^c \in 10_f$ and $N^c \sim 1_f$. In order to introduce
$\delta^+$, one needs to add to the model a 10-dimensional Higgs multiplet, which
has the proper couplings required in section 2 to achieve RH leptogenesis:  $Y_R 1_f
\overline{10}_f 10_H$ and $Y_L \overline{5}_f  \overline{5}_f 10_H$.

At the end of section 2 we discussed possible ``Zee-like" trilinear couplings between
$\delta^+$ and two different Higgs doublets $H_i$ with hypercharge -1. 
In the left-right symmetric models, only one $H_i$ is contained in the bidoublet
field (2,2,0) which provides the usual Dirac-type Yukawa couplings. In Pati-Salam
models, there are two such doublets, one in (2,2,1) and one in (2,2,15), but
$SU(4)_c$ invariance prevents them to couple to $\delta^+ \in (1,1,10)$. Similarly,
in $SO(10)$ context, $10_H,~\overline{126}_H,~120_H$, contain, respectively, one, one and two fields $H_i$, but $\delta^+\in120$ has no trilinear coupling to them. Therefore,
the ``Zee-like" coupling requires the introduction of at least one extra Higgs multiplet
with no Yukawa couplings to fermions, e.g., $210_{SO(10)} \supset
(2,2,\overline{10})_{SU_{224}} \supset (2,2,-2)_{SU_{221}}$.
In $SU(5)$ models, one $H_i$ is contained either in $\bar{5}$ or in $\overline{45}$ and
any choice of two such doublets is sufficient to couple to $\delta^+\in 10$.

Even in models where the couplings $\delta^+ H_i H_j$ are present, they become dangerous for leptogenesis only if both Higgs doublets are lighter than $M_{N_1}$, which generically requires fine-tuning. In this case
these couplings must be suppressed
or alternatively the coupling $Y_L$ must be
forbidden. This last option may be naturally realized
requiring that lepton number is broken only softly, by RH neutrino 
Majorana masses.

\subsection{The case of a right-handed triplet $\Delta_R$ \label{MB2}}

In the case of leptogenesis via a RH triplet $\Delta_R =
(\delta^{++},\delta^+,\delta^0)$, it is understood that
the gauge symmetry includes, at least, the minimal left-right symmetric group
$SU(2)_L \times SU(2)_R \times U(1)_{B-L}$.
A RH triplet is naturally present in left-right models
since the VEV of its neutral component $\delta^0$ provides the correct symmetry
breaking to the Standard Model and, moreover, it gives a Majorana mass to the 
RH neutrinos. In fact, $\Delta_R \sim (1,3,2)$ couples symmetrically to two RH
lepton doublets $\psi_R \sim (1,2,-1)$. In Pati-Salam models, $\Delta_R$ is
contained in the $(1,3,10)$ multiplet which, in turn, belongs 
to $\overline{126}$
Higgs representation in $SO(10)$.

Let us discuss limits and merits of the minimal left-right symmetric framework,
in order to realize the RH leptogenesis scenario we proposed in 
section 3.\footnote{For more standard realizations of leptogenesis 
in the minimal left-right model based on RH neutrino decay to 
left-handed leptons 
or on $SU(2)_L$ triplet scalar decay, see Refs.~\cite{hs,lrlepto}.}
The mass of the $\delta^+$ can be smaller than 
the $SU(2)_R$ breaking scale (as well as the RH neutrino masses)
since they are determined by independent scalar potential couplings.
The left-right model has the nice feature that $\delta^+$ undergoes 
only three-body
decays, since $\Delta_R$  does not couple to left-handed lepton doublets nor
to two Higgs bosons, as long as $SU(2)_L$
is unbroken (these last couplings can come only from 
the interaction $\lambda_{ij} 
Tr (\Delta_R^\dagger \Phi_i \Delta_L \Phi_j^\dagger)$,
where $\Phi_i$ are bidoublet Higgs bosons). These couplings if present 
would induce fast $\Delta L = 2$ scattering washout effects.
The presence of the left-handed triplet $\Delta_L$ might also induce 
washout effects, which are suppressed, however, if $\Delta_L$
is heavier or has small $\lambda_{ij}$  
couplings.
The effect of the RH gauge bosons are suppressed because 
they have a mass naturally of the order of or 
heavier than the heaviest RH neutrino.

Notice that, in section 3, we introduced a Majorana mass 
term $M_R$ for RH neutrinos in addition to the Yukawa 
coupling $Y_\Delta$ between $\Delta_R$ and RH lepton doublets.
In the minimal left-right model, the two terms can be 
identified, so that only one 
source of flavor breaking is present:
the matrices $M_R$ and $Y_\Delta$ are 
proportional to each other. As a consequence, in the $N_i$ mass 
eigenstate basis, 
the $Y_\Delta$ coupling matrix is also diagonal and the diagrams of Fig.~1 
with two different RH neutrinos are simply vanishing.
Therefore, for this leptogenesis mechanism to be effective we 
need to extend the minimal model in order to distinguish $M_R$ 
from $Y_\Delta$. For example, one may introduce 
a second RH triplet (a second $\overline{126}$ in $SO(10)$),
or consider extra (e.g.~non-renormalizable) sources of RH neutrino mass.
Alternatively, one could resort to the singlet leptogenesis 
mechanism, adding a $(1,1,2)$ Higgs multiplet ($120$ in $SO(10)$ context).  

The dangerous couplings between $\delta^+$ and two Higgs 
doublets $H_i$, absent in the minimal left-right model, 
could appear in more general theories. The discussion 
is completely parallel to the singlet case of the previous 
section: such couplings are forbidden as long as one 
considers only $H_i$ contained in Higgs multiplets with Yukawa 
couplings to fermions. However they appear, for example, if a $210$ 
Higgs is introduced in $SO(10)$ models (and similarly for 
smaller gauge groups). 

Note finally that in Grand-Unified models, although one generally expects 
the right-handed neutrinos as well as the scalar singlet $\delta^+$
to be very massive, 
it is not excluded that these particles are present at the TeV scale. One may 
worry in this case that
the charged singlet scalar (at the intermediate scale) affects the running
of the $U(1)_Y$ gauge coupling (with a contribution $\Delta b_1 = 1/5$
to the $\beta$-function, in the usual normalization).
However, the full particle spectrum at the intermediate scale
is highly model-dependent. In particular, the scale of the
left-right symmetry breaking $v_R$ can be as small as a few TeV and be 
consistent with unification (for a recent analysis see Ref.~\cite{lowlr}).


\section{Other possible leptogenesis contributions from the RH sector}

In this section we identify other minimal mechanisms to induce a 
lepton asymmetry from the RH lepton sector.

One possibility is 
to have two RH neutrinos quasi-degenerate in mass. In this case the 
self-energy diagram of Fig.~1(b) leads to resonant enhancement of the asymmetry
pretty much as in the ordinary leptogenesis model.
One may expect that this enhancement allows to achieve successful 
leptogenesis at low scale relaxing the hierarchy between the 
couplings $(Y_R)_{ij}$ discussed in section \ref{singlet}. However, in 
order to have observable consequences, 
in addition to requiring strongly degenerate RH neutrino masses, 
this scenario still needs a hierarchy of couplings \cite{pil2},
similar to (or slightly smaller than) the one considered 
above (i.e. small $N_1$ couplings in order to 
avoid too large washout from inverse decays and larger $N_{2,3}$ couplings 
to lead to direct observations of RH neutrinos). Only very close 
to the resonance
larger $(Y_R)_{1j}$ couplings (up to few $10^{-5}$) can be taken 
for $M_{N_1}\sim$ few TeV. The improvement with respect to the resonant case
in the usual leptogenesis model is that
the Yukawa couplings leading to observable consequences do not 
require cancellation between them to avoid the generation of too large 
neutrino masses.

Another possibility of successful leptogenesis we 
want to mention, and which could work in the minimal 
left-right model or SO(10), 
occurs at high scale if the $\delta^+$ is lighter than
the RH neutrinos. In this case, the lepton number violating 
3-body decay of the $\delta^+$, 
Eq.~(\ref{deltawidth2}), can induce leptogenesis considering 
the ordinary vertex and self-energy 
one loop diagrams of the virtual $N_i$ in this decay.
This leads to asymmetries of the
order $\varepsilon_{N_1}\cdot (M_\delta/M_{N_1})^2$ (where 
$\varepsilon_{N_1}$ is the ordinary two body decay 
asymmetry of $N_1$ decaying to 
left-handed leptons). As a result, and taking into account the fact 
that the $\delta^+$ can be thermalized by gauge scatterings, we estimate
that this 3-body decay can lead to successful leptogenesis 
for $M_\delta \gtrsim 10^{11}$ GeV.
Note that, due to the fact that the 3-body decay 
of the $\delta^+$ is quite slow,  in general it will not
washout preexisting lepton asymmetries, in particular the 
asymmetry which could have been produced by 
the decay of the $N_i$ at a higher scale.

Finally, we consider other possible mechanisms of successful 
leptogenesis driven by the other components of an $SU(2)_R$ 
triplet $\Delta_R$, that is $\delta^{0}$ and $\delta^{++}$. The 
neutral component $\delta^0$, if lighter than RH neutrinos $N_i$, 
undergoes 4-body decay into two left-handed leptons and 
two Higgs doublets, via two 
virtual $N_i$. The lepton asymmetry generated by these decays comes from 
the ordinary one-loop 
diagrams of both virtual $N_i$. It is therefore proportional 
to $\sim \epsilon_{N_i}$ but it is suppressed by extra powers 
of $M_{N_i}/M_{\delta^0}$. The 
decay width is extremely suppressed and therefore satisfies easily the 
out-of-equilibrium condition. We estimate that leptogenesis can be 
successful only at very high scale, $M_{\delta^0}\gtrsim 10^{11}$ GeV. 

The $\delta^{++}$ decays into two RH charged leptons. Such a decay can produce
a lepton 
asymmetry only if there are at least two 
different $\delta^{++}$, via 
the self-energy diagram involving two charged Higgs 
bosons (just like with two left-handed 
triplets \cite{ms,hms}). The extra scalars in the loop are 
naturally given by the $\delta^+$, since the scalar potential
term $Tr(\Delta_R\Delta_R)Tr(\Delta_R^\dag \Delta_R^\dag)$ provides, 
after $SU(2)_R$ breaking, the trilinear 
coupling $\delta^{++}\delta^-\delta^- + h.c.$.
This mechanism requires triplets with mass above $\sim 10^{10}$ GeV (except 
if they are quasi-degenerate).
This model does not present any particular advantage with respect to the more 
straightforward type I and/or II seesaw models 
of leptogenesis. It illustrates once more the fact that there 
are many possible leptogenesis models at a 
high scale, but only very few working at low scale.

   
\section{Phenomenology of a TeV scale charged $SU(2)_L$ singlet  scalar}

The observation of a light $SU(2)_L$ singlet $\delta^+$ at 
colliders  would 
imply that, in the presence of RH neutrinos, the $Y_R$ 
interactions occur naturally.
This would render our leptogenesis 
mechanism as plausible as the standard one. 
Moreover the fact that this 
model can work at scales 
as low as the TeV scale opens the possibility to produce 
directly a 
RH neutrino through the relatively  large $Y_R$ couplings 
of the $N_2$ and/or $N_3$, which can have a mass as low as 
few TeV.\footnote{The production of TeV scale RH neutrinos through 
the Yukawa couplings to left-handed leptons has been 
discussed e.g. in \cite{pp} for LHC, \cite{AG} for a high energy $e^+e^-$ linear collider and
\cite{piga} for an $e\gamma$ collider.} 
This would leave in general no 
other choice for leptogenesis (and baryogenesis) than to be produced at low 
scale below $M_{N_{2,3}}$, as allowed by our 
model.\footnote{A possible exception is the 
case where the observed $N_i$ has suppressed 
couplings to a given flavor, so that it cannot 
washout any preexisting lepton asymmetry associated to that flavor.}

Note that to produce a $\delta^+$ at colliders, the 
Drell-Yan $\delta^+ \delta^-$ pair 
production process (from $e^+ e^-$ or $q \bar{q}$ annihilation with 
an intermediate
photon or $Z$) is the most effective way.
For definiteness, consider the differential cross-section for
$e^+e^- \rightarrow \gamma \rightarrow \delta^+\delta^-$:
$$
\frac {d\sigma} {d\cos\theta} = \frac {\pi\alpha^2}{4Q^2}
\left(1-\frac{4M^2_\delta} {Q^2}\right)^{3/2} \sin^2\theta ~,
$$
where $Q^2$ is the center-of-mass energy squared, $\alpha$ is the
electromagnetic constant
and $\theta$ the angle between the collision axis and the outgoing
$\delta^+$. For $Q^2=500$ GeV (as foreseen at the International Linear 
Collider) a $\delta^+$ of about 200 GeV would be produced with a total cross
section of about 20 femtobarns.
For any scalar with given weak isospin and hypercharge, the 
pair production cross section from $q \bar{q}$ annihilation can be found 
e.g.~in  \cite{Muhl}.  In particular, this paper studies 
the $pp \rightarrow \Delta^{++} \Delta^{--}$ cross section relevant 
for LHC with $\Delta^{++}$ a doubly charged particle member of a $SU(2)_L$ 
scalar triplet. This cross section has been found to be large 
enough by far to observe a $\Delta^{++}$ with a mass as large as 1 TeV.
The LHC $pp \rightarrow \delta^+ \delta^-$ 
pair production 
is similar, up to factors of order unity due to different charges. 

The identification of $\delta^+$ relies on the comparison of its dominant
decay channels with background, which is beyond the scope of this article. 
Just note that, as we discussed above, if $\delta^+$
couples (antisymmetrically) to left-handed leptons as in Section 2, 
it should mainly decay
into antilepton and antineutrino. The decay into anti-$\tau$ 
is actually the more
interesting one, because it can be used to identify at the LHC 
a relatively light MSSM charged Higgs $H^+$
(see, e.g., \cite{hashemi}).  The two particle decays can
be distinguished by analyzing the angular distribution of the outgoing
antilepton, since it is left-handed in the case of $H^+$ and right-handed
in the case of $\delta^+$. 
In case the $Y_L$ couplings would be in addition 
suppressed (below $\sim 10^{-7}$),
the lifetime of the $\delta^+$ would be much longer than for $H^+$, leading
to a displaced vertex when it decays.

Note also that the $\delta^+$ singlet 
can induce, through its $Y_L$ couplings,
a $\mu \rightarrow e \gamma$ transition with branching ratio
Br$(\mu \rightarrow e \gamma)\approx
(\alpha/48\pi)|(Y_L)_{e \tau} (Y_L)_{\mu \tau}|^2/$ $(M_\delta^4 G_F^2)$ (see
e.g.~\cite{his}). With $M_\delta$ below TeV, a branching ratio of the order of
the experimental limit (Br$(\mu \rightarrow e \gamma)< 1.2 \cdot 10^{-11}$ 
at $90\%$ C.L. \cite{meg}) can 
be easily obtained. Similarly the $Y_R$ couplings can induce this transition
with Br$(\mu \rightarrow e \gamma)\approx
(\alpha/192\pi)|(Y_R)_{i e} (Y_R)_{i\mu}|^2/$ $(M_{N_i}^4 G_F^2)$, where 
we assumed that the exchange of the RH neutrino $N_i$ gives the main 
contribution and we neglected $M_\delta/M_{N_i}$ corrections.
In this case the sets of parameters which lead to successful leptogenesis 
give rise  
to a smaller branching ratio, below $\sim 10^{-17}$, therefore unobservable.

The case of the triplet $(\delta^0, \delta^+, \delta^{++})$ has a 
similar phenomenology for what concerns the production of the $\delta^+$
and $N_{2,3}$. However, here $\delta^+$ does not have 2-body decays 
since it has no $Y_L$ coupling and it can 
decay only very slowly to three bodies (see Section 3).
Therefore,  the Drell-Yan produced $\delta^+ \delta^-$ pair will leave 
in the detector a pair of long curved charged particle tracks which could 
be distinguished from a muon pair by the fact that they would be 
less relativistic. In this case the decay to a charged lepton pair and a $H^+$
(see Eq.~(\ref{deltawidth1})) will occur in general outside the 
detector and cannot be seen.
In this scenario a $\delta^{++}$ could also be produced 
electromagnetically 
in colliders. As there is no $Y_L$ couplings, the $\mu \rightarrow e \gamma$
process in this case can be induced only through the $Y_\Delta$ couplings,
with 
suppressed branching ratios as for the singlet case with $Y_R$ couplings.


\section{Summary}

We have considered a new mechanism to induce leptogenesis 
successfully, by the decay of the RH neutrino $N_1$ 
to a RH charged lepton 
and a scalar $SU(2)_L$ singlet $\delta^+$.
In the presence of left-right symmetry
the $\delta^+$ may or may not be
a member of an $SU(2)_R$ triplet.
In both versions one achieves successful leptogenesis easily in a similar way.
This mechanism can work at scales 
as low as few TeV with no need of resonant enhancement of the asymmetry. 
Such a low scale realization  
requires that $N_1$ Yukawa couplings to RH charged leptons 
are about 4 orders of magnitude smaller than the ones of heavier RH neutrinos.

In grand-unified theories this mechanism
can be realized,  for the singlet case, 
both in SO(10), if there exists a 120 scalar multiplet, and in 
SU(5) with a 10 scalar multiplet. 
The $SU(2)_R$ scalar triplet case can be 
incorporated in SO(10) models with a $\overline{126}$ scalar multiplet. 
However, in this case, in order for leptogenesis to work, the model should 
contain a source of RH neutrino 
masses independent from this $\overline{126}$ representation.

Phenomenologically, the observation of a light $SU(2)_L$ singlet $\delta^+$ at 
colliders would be a strong evidence in favor of our proposal.  The additional 
production of a RH neutrino at few TeV scale,
through the large couplings to RH charged leptons,  
would make the case for low scale leptogenesis.

\section*{Acknowledgments}

We thank Louis Fayard for useful discussions. 
MF is partially supported by the RTN European Program MRTN-CT-2004-503369.
The work of TH is supported by a Ramon y Cajal contract of the Ministerio de 
Educaci\'on y Ciencia. 
EM is supported in part by the U.S. Department of Energy under Grant 
No.~DE-FG03-94ER40837.
TH thanks the Service de Physique Th\'{e}orique, CEA-Saclay, for hospitality during the initial stage of this project.


\end{document}